\documentclass[9pt]{article}
\usepackage{amsmath}
\usepackage{graphicx}
\usepackage[margin=1in,footskip=0.5in]{geometry}
\usepackage{amssymb}
\usepackage{caption}
\usepackage{subcaption}
\usepackage{epstopdf}
\usepackage[english]{babel}
\usepackage{graphicx}
\usepackage{rotating}
\usepackage{caption}
\usepackage[autostyle]{csquotes}
\usepackage[titletoc,toc,title]{appendix}
\usepackage{varwidth}
\usepackage{circuitikz}
\usepackage{enumitem}
\setlist{nolistsep}
\usepackage[justification=justified]{caption}
\usepackage{authblk}

\begin{document}

\title{ Manipulation of entanglement sudden death in an all-optical setup}


\author[1]{Ashutosh Singh}
\author[1,2]{Siva Pradyumna}
\author[3]{A.R.P. Rau}
\author[1*]{Urbasi Sinha}
\affil[1]{Light and Matter Physics Group, Raman Research Institute, Sadashivanagar, Bangalore 560080, India.}
\affil[2]{Politecnico di Torino, Istituto Nazionale di Ricerca Metrologica, 10135 Torino TO, Italy.}
\affil[3]{Department of Physics and Astronomy, Louisiana State University - Baton Rouge, LA 70803, USA.}
\affil[*]{Corresponding author: usinha@rri.res.in}


\date{}
\maketitle

\begin{abstract}

The unavoidable and irreversible interaction between an entangled quantum system and its environment causes decoherence of the individual qubits as well as degradation of the entanglement between them. Entanglement sudden death (ESD) is the phenomenon wherein disentanglement happens in finite time even when individual qubits decohere only asymptotically in time due to noise. Prolonging the entanglement is essential for the practical realization of entanglement-based quantum information and computation protocols. For this purpose, the local NOT operation in the computational basis on one or both qubits has been proposed. Here, we formulate an all-optical experimental set-up involving such NOT operations that can hasten, delay, or completely avert ESD, all depending on when it is applied during the process of decoherence. Analytical expressions for these are derived in terms of parameters of the initial state's density matrix, whether for pure or mixed entangled states. After a discussion of the schematics of the experiment, the problem is theoretically analyzed, and simulation results of such manipulations of ESD are presented.
\end{abstract}



\maketitle

\section{Introduction}

Quantum entanglement [1,2] is a non-classical correlation shared among quantum systems which could be non-local [3,4] in some cases. It is a fundamental trait of quantum mechanics. Like classical correlations, entanglement also decays with time in the presence of noise in the ambient environment. The decay of entanglement depends on the initial state and the type and amount of noise (Amplitude damping, Phase damping, etc.) acting on the system [5-7]. The entangled states: $|\psi^\pm\rangle=|\alpha| |ge\rangle\pm|\beta| \exp{(\iota \delta)} |eg\rangle)$ and $|\phi^\pm\rangle=|\alpha| |gg\rangle\pm|\beta|  \exp{(\iota \delta)} |ee\rangle$ (maximally entangled ``Bell states`` for $|\alpha|=|\beta|=1/\sqrt{2},~\delta=0$) being the simplest and most useful entangled states in quantum information processing receive special attention. The maximally entangled states $|\phi^\pm\rangle$ and $|\psi^\pm\rangle$ undergo asymptotic decay of entanglement in the presence of an amplitude damping channel (ADC). The non maximally entangled states $|\psi^\pm\rangle$ always undergo asymptotic decay of entanglement, whereas $|\phi^\pm\rangle$ undergo asymptotic decay for $|\alpha|>|\beta|$ and a finite time end called entanglement sudden death (ESD) for $|\alpha|<|\beta|$ in the presence of ADC. On the other hand, a pure phase damping channel (PDC) causes entanglement to always decay asymptotically. Two different initial states ($|\psi\rangle=|\alpha|~|gg\rangle+|\beta| \exp{(\iota \delta)}~ |ee\rangle, ~|\alpha|^2+|\beta|^2=1~$, where $ ~ (i)~|\beta|=k |\alpha|~~\&~~(ii) ~|\alpha|=k|\beta|,~k>1~$), which share the same amount of initial entanglement (measured through Negativity) being affected by the same type of noise may follow very different trajectories of entanglement decay. In the presence of multiple stochastic noises, although the decoherence of individual qubits follows the additive law of relaxation rates, the decay of entanglement, does not. In fact, entanglement may not decay asymptotically at all, and disentanglement can happen in finite time (ESD). ESD has been experimentally demonstrated in atomic [8] and photonic systems [9,10].

Since entanglement is a resource in  quantum information processing [11-13], manipulation that prolongs entanglement will help realize protocols that would otherwise suffer due to short entanglement times. Also, entanglement purification [14] and distillation [15] schemes could possibly recover the initial correlation from the ensemble of noise-degraded correlation so long as the system has not completely disentangled. Therefore, the delay or avoidance of ESD is important. Several proposals exist to suppress the decoherence; for example, decoherence-free subspaces [16-19], quantum error correction [20,21], dynamical decoupling [22-24], quantum Zeno effect [25-27], quantum measurement reversal [28-33], and delayed-choice decoherence suppression [34]. Protecting entanglement using weak measurement and quantum measurement reversal [32,33], and delayed choice decoherence suppression [34] have been experimentally demonstrated. Both of these schemes, however, have the limitation that the success probability of decoherence suppression decreases as the strength of the weak interaction increases.

The practical question we want to address here is; whether, given a two-qubit entangled state in the presence of amplitude damping channel which causes disentanglement in finite time, can we alter the time of disentanglement by a suitable operation during the process of decoherence?  A theoretical proposal exists in the literature for such manipulation of ESD [35] through a local unitary operation (NOT operation in computational basis: $\sigma_x$) performed on the individual qubits which swaps their population of ground and excited states. Depending on the time of application of this NOT operation, it can avoid, delay, or hasten the ESD. Based on this proposal [35], we have extended the experimental set up [9] for ESD and propose here an all-optical experimental set up for manipulating ESD involving the NOT operation on one or both the qubits of a bipartite entangled state in a photonic system.

The system consists of polarization-entangled photons ($|\psi\rangle=|\alpha||HH\rangle + |\beta|  \exp{(\iota \delta)} |VV\rangle$) produced in the sandwich configuration Type-I SPDC (spontaneous parametric down conversion)[38]. These photons are sent to two displaced identical Sagnac interferometers, where ADC is simulated using rotating HWPs (half wave plates) placed in the path of incoming photons (See Fig.2). The HWP selectively causes a $|V\rangle$ polarized photon to ``decay`` to $|H\rangle$ ($|H\rangle$ and $|V\rangle$ serve as ground and excited states of the system, the two states of a qubit)[9]. The NOT operation is implemented by a HWP with fast axis at $45^o$ relative to $|V\rangle$, placed right after the ADC. This HWP  is followed by PBS (polarizing beam splitter) to segregate the $|H\rangle$ and $|V\rangle$ polarizations and, with subsequent ADC after the NOT operation implemented by a set of secondary  HWPs acting on $|V\rangle$ only. Such a set of secondary HWPs simulating the ADC (or evolution of qubits in noisy environment) is essential to our study as the ADC (for example, spontaneous emission in case of a two-level atomic system) continues to act even after the NOT operation is applied [35] and these secondary HWPs simulate it in our proposed experiment. The orientation of the HWP ($\theta$) plays the role of time ($t$) with $\theta\rightarrow 45^o$ ($p=\sin^2(2\theta)\rightarrow 1$) analogous to $t\rightarrow \infty$ ($p=1-\exp{(-\Gamma t)}\rightarrow 1$) for a two-level atomic system decaying to the ground state due to spontaneous emission.

We use Negativity as a measure of entanglement. It is defined as the sum of absolute values of negative eigen values of the partially transposed density matrix [36,37]. We find that our simulation results for the manipulation of ESD involving NOT operations on one or both the qubits of a polarization entangled photonic system in presence of ADC are completely consistent with the theoretical predictions of the reference [35] which has analyzed an atomic system. The merit of our scheme is that it can delay or avoid ESD (provided the NOT operation is performed sufficiently early) unlike previous experiments [32-34] where success probabilities scaled with the strength of the weak interaction. Since the photonic system is time independent and noise is simulated using HWPs, it gives experimentalists complete  freedom to  study and manipulate the disentanglement dynamics in a controlled manner. In this, our photonic system through a controllable HWP offers an advantage over others such as atomic states where the decay occurs through noise sources lying outside experimental control. The NOT operation that we apply through a HWP is the analogue of flipping spin in a nuclear magnetic system, achieved through what is referred to as a $\pi$ pulse.

The paper is organized as follows: In section (2), we discuss the all-optical implementation of the proposed ESD experiment and analyze it theoretically using the Kraus operator formalism. In section (3) and (4), we discuss and theoretically analyze the proposed ESD-manipulation experiment involving NOT operation on both or on only one of the qubits, respectively. In section (5), we give analytical expressions for probabilities $p_0,p_A,~\&~p_B$ and also for ESD and its manipulation curves in terms of the parameters of the initial state (density matrix). The first of these is the setting (``time``) for ESD, the next setting for the NOT marking the  border between hastening and delay; that is, if the NOT is applied after $p_A$ (and of course before $p_0$), it actually hastens, ESD happening before $p_0$, whereas application before delays ESD to stretch past $p_0$ to larger but still finite value less than one. The third, $p_B$, marks the border between delaying or completely avoiding ESD. Applying the NOT after $p_B$ delays to a larger $p_0$ value whereas applying before avoids ESD altogether. In section (6), we summarize the results of manipulation of ESD for different pure and mixed initial entangled states, giving  numerical values of $p_0,p_A$, and $p_B$. Section (7) concludes with pros and cons of the proposed scheme for the manipulation of ESD and the future scope of this work.

\section{Proposed experimental set up for ESD and its analysis} 

The proposed experimental setup for ESD is shown in figure (1), which is a generalization of the scheme used in reference [9]. The type-1 polarization entangled photons ($|\psi\rangle=|\alpha||HH\rangle+|\beta|~  \exp{(\iota\delta)}|VV\rangle$) can be prepared by standard methods [38]; the amplitudes $|\alpha|$ and $|\beta|$ and relative phase $\delta$  are controlled by the HWP and QWP (quarter wave plate). These entangled photons are sent to two displaced Sagnac interferometers with HWPs simulating the ADC, where decoherence takes place, and finally these photons are sent for tomographic reconstruction of the quantum state [39]. The $|H\rangle$ and $|V\rangle$ polarizations of the photon serve as the ground and excited states of the analogous atomic system, while output spatial modes of the reservoir $|a\rangle,~|a'\rangle~~\&~~|b\rangle,~|b'\rangle$ serve as the ground and excited states of the reservoir. Asymmetry between degenerate polarization states of a photon ( $|H\rangle~\&~|V\rangle$) is introduced by the HWP rotation such that it selectively causes an incident $|V\rangle$ polarization to ``decay`` to $|H\rangle$, while leaving $|H\rangle$ polarization intact. Thus the $|H\rangle$ and $|V\rangle$ polarization states are analogous to the ground and excited states, respectively, of a two-level atomic system.

\begin{figure}
\begin{center}
\includegraphics[clip, trim=3.5cm 9cm 0.5cm 2cm, width=0.55\textwidth]{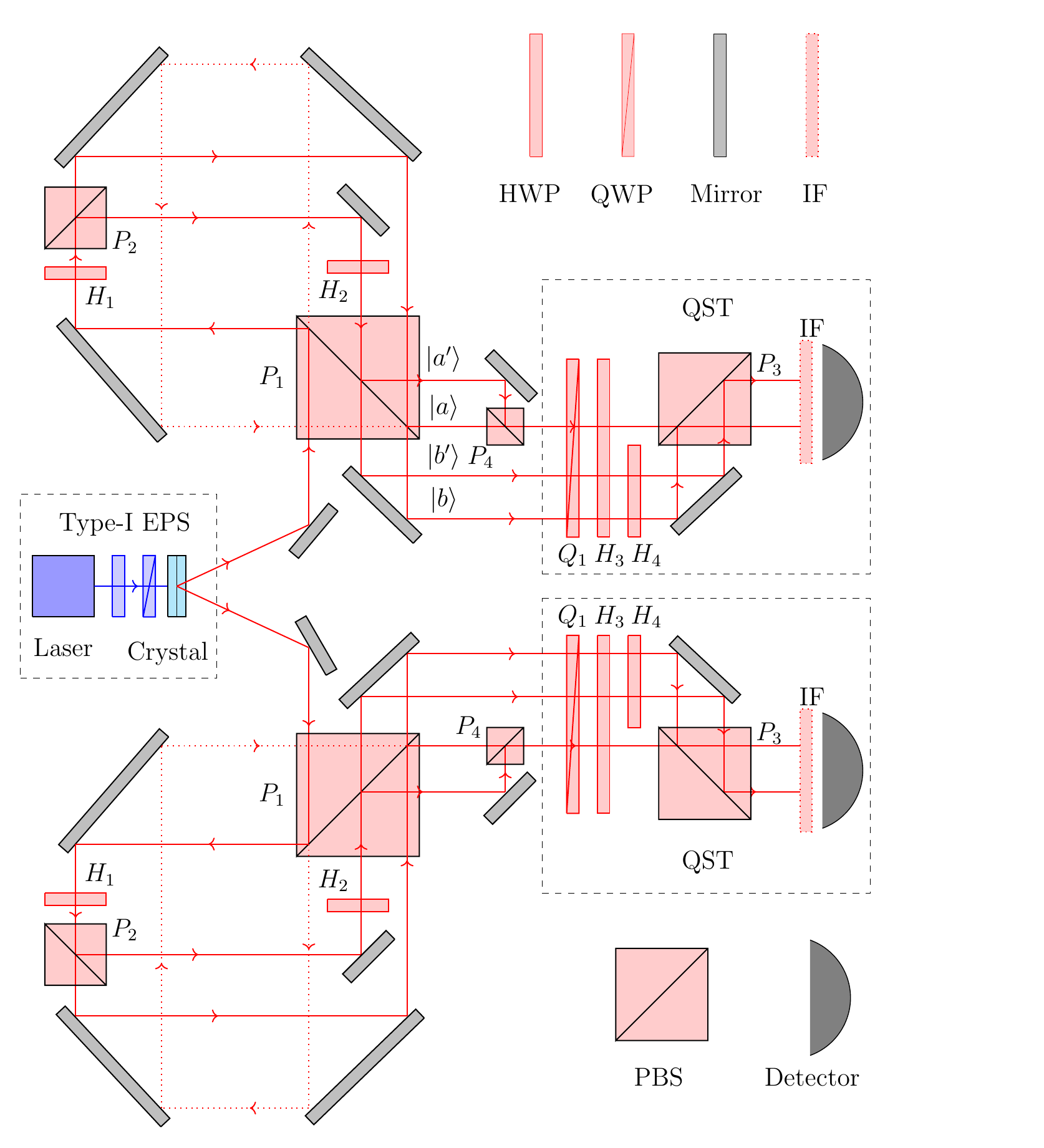}
\caption{\textit{The proposed experimental set up for ESD in the presence of ADC in a photonic system. The polarization entangled photons from the mid-left of the figure are sent to two displaced identical Sagnac interferometers. The figure shows the Sagnac interferometers and its output spatial modes: $|a\rangle,|a'\rangle, |b\rangle ~\&~ |b'\rangle$, which serve as the modes of the reservoir. Following an initial photon in the upper arm, an incident $|V\rangle$ polarized photon is reflected by PBS $P_1$ and traverses the interferometer in clockwise sense where HWP $H_1$ acts as ADC causing $|V\rangle$ polarized photon to decay to $cos(2\theta)|V\rangle+sin(2\theta)|H\rangle$. The PBS $P_2$ transmits $|H\rangle$ component and reflects $|V\rangle$ component. The transmitted $|H\rangle$ component comes back to PBS $P_1$ and is transmitted to spatial mode $|b\rangle$. The reflected $|V\rangle$ component from $P_2$ sees another ADC HWP $H_2$ which causes it to decay to $cos(2\theta')|V\rangle+sin(2\theta')|H\rangle$ and when it comes to PBS $P_1$, the $|V\rangle$ component gets reflected to mode $|a'\rangle$ and $|H\rangle$ component gets transmitted to mode $|b'\rangle$. On the other hand, an incident $|H\rangle$ polarized photon is transmitted by the PBS $P_1$ and traverses the interferometer in counter-clockwise sense, comes back to PBS $P_1$ and gets transmitted to mode $|a\rangle$. The path lengths of the photons reaching in modes $|a'\rangle$ and $|a\rangle$ are compensated for coherent recombination of polarization amplitudes at PBS $P_4$. This ensures that when all the ADC HWPs are set to zero, and an initial entangled state $|\alpha||HH\rangle+|\beta|\exp(\iota\delta)|VV\rangle$ is incident at the input ports of the interferometers, the initial state is reconstructed at the output ports. These photons are finally sent for quantum state tomography (QST). The HWP $H_4$ is used to flip the polarization of photons passing through it such that the QST settings remain the same for the photons in all spatial modes. The Q,H and P stand for Quarter wave plate, Half wave plate and Polarizing beam splitter. EPS and IF stand for Entangled Photon Source and Interference Filter respectively. For tracing photon paths with polarizations and spatial modes, see also Appendix [A].}}
\end{center}
\end{figure}

The ADC is implemented using two HWPs: $H_1~\&~H_2$ oriented at $\theta~\&~\theta'$ respectively, such that incident $|V\rangle$ polarization amplitude ``decays`` to $|H\rangle$. For different fixed orientations of $H_1$, evolution in ADC is completed by rotating $H_2$. The PBS $P_2$ is used to segregate the $|H\rangle$ and  $|V\rangle$ polarization amplitudes such that the HWP $H_2$ is applied only to $|V\rangle$ polarization for it to serve as excited state of the system and leaving $|H\rangle$ polarization (ground state of the system) undisturbed. 

The single qubit Kraus operators for ADC are given by,

\begin{equation}
M_1= \left( \begin{array}{cccc} 1 & 0\\
0 & \sqrt{1-p} \end{array} \right)~~,~~
M_2=\left( \begin{array}{cccc} 0 &\sqrt{p}\\
0 & 0  \end{array} \right),
\end{equation}
where $p=\sin^2(2\theta)$ for ADC mimicked by HWP in a photonic system [9].

These operators satisfy the completeness condition,
\begin{equation}
M_1^\dag M_1+M_2^\dag M_2=\mathbb{I} ,
\end{equation}
where $\mathbb{I}$ is the identity matrix.

The Kraus operators for the two qubits are obtained by taking appropriate tensor products of single qubit Kraus operators as follows, 

\begin{equation}
M_{ij}=M_i \otimes M_j~~;~~i,j=1,2 .
\end{equation}

Label another set of Kraus operators by $M'_{ij}~;~i,j=1,2$, with variable $p$ replaced by $p'$ ($p'=\sin^2(2\theta')$) to distinguish it from the former, with the form of Kraus operators remaining similar to that in (1). Such a splitting into two angles or two values of probability will prove convenient for later applications in the section of manipulation using optical elements in between. 

Let the initial state of the system be
\begin{equation}
|\psi\rangle=|\alpha| |HH\rangle+|\beta| \exp(\iota\delta)|VV\rangle ,
\end{equation}
with a corresponding density matrix given by,
\begin{equation}
\rho(0,0)= \left( \begin{array}{cccc} 
u & 0 & 0 & v\\
0 & 0 & 0 & 0 \\
0 & 0 & 0 & 0 \\
v^* & 0 & 0  & x 
\end{array} \right),
\end{equation}
where $u=|\alpha|^2~,~v=\alpha\beta^*~,~v^*=\beta\alpha^*~,~x=|\beta|^2$ and $u+x=1$. In general, if $|v|^2=ux$, this represents a pure entangled state, otherwise a mixed entangled state. A more general mixed state with non-zero entries in the other two diagonal positions is considered in appendix [B].   

The initial state of the system (5) in the presence of ADC (due to $H_1$ at $\theta$)  evolves as follows,

\begin{equation}
\rho^{(1)}(p,0)=\sum _{i,j} M_{ij}~\rho(0,0)~M_{ij}^\dag~~;~~i,j=1,2.
\end{equation} 

Apply the Kraus operators $M'_{ij}$ to complete the evolution in the presence of ADC (due to $H_2$ at $\theta'$) as follows,

\begin{equation}
\begin{aligned}
&\rho^{(1)}(p,p')=\sum _{i,j} M'_{ij}~\rho(p,0)~M_{ij}^{\prime\dag}~~;~~i,j=1,2~, \\
&=\left(\begin{array}{cccc}
 \rho^{(1)}_{11}(p,p') & 0 & 0 &  \rho^{(1)}_{14}(p,p') \\
 0 &  \rho^{(1)}_{22}(p,p') & 0 & 0 \\
 0 & 0 &  \rho^{(1)}_{33}(p,p') & 0 \\
  \rho^{(1)}_{41}(p,p')& 0 & 0 &  \rho^{(1)}_{44}(p,p') \\
\end{array}\right),
\end{aligned}
\end{equation} 

where,

\begin{equation}
\begin{aligned}
\rho^{(1)}_{11}(p,p')&=u + p^2 x+ p'^2 (1 - p)^2 x + 2 p' (1 - p) p x ,\\
\rho^{(1)}_{22}(p,p')&=(1 - p') p' (1 - p)^2 x + (1 - p') (1 - p) p x,\\
\rho^{(1)}_{33}(p,p')&=(1 - p') p' (1 - p)^2 x + (1 - p') (1 - p) p x,\\
\rho^{(1)}_{44}(p,p')&=(1 - p')^2 (1 - p)^2 x,\\
\rho^{(1)}_{14}(p,p')&=(1 - p') (1 - p) v,\\
\rho^{(1)}_{41}(p,p')&=(1 - p') (1 - p) v^*.\\    
\end{aligned}
\end{equation}
 
 \section{Proposed experimental set up for manipulation of ESD using the NOT operation on both qubits of a bipartite entangled state}

The proposed experimental set up for manipulation of ESD based on the local NOT operation performed on both the qubits of a bipartite entangled state (5) is shown in figure (2). The HWP $H_1$ acts as ADC for incident $|V\rangle$ polarized photon and then NOT operation is performed by $H_5$ at $45^o$, which swaps the $|H\rangle$ and $|V\rangle$ amplitudes, which are then segregated by PBS $P_2$. The ADC is continued by synchronous rotation of $H_2~\&~H_6$ oriented at $\theta'$, which causes the swapped $|V\rangle$ amplitude to ``decay`` to $|H\rangle$. The photons from the output spatial modes of the interferometer are sent for tomographic reconstruction of the quantum state [39].

\begin{figure}[htb]
\begin{center}
\includegraphics[clip, trim=3.5cm 9cm 2.5cm 2cm, width=0.5 \textwidth]{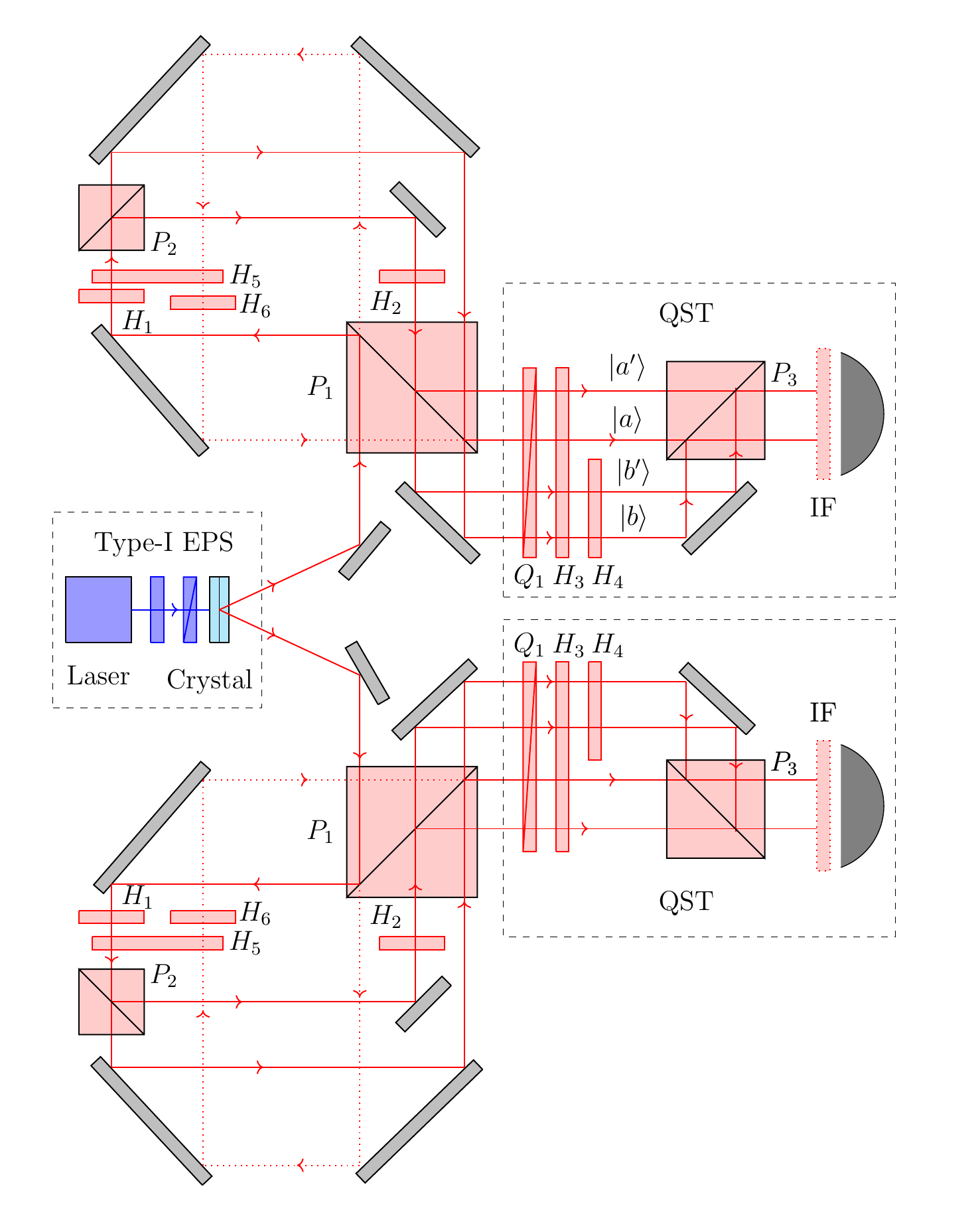}
\caption{\textit{The proposed experimental set up for manipulation of ESD involving NOT operation on both the qubits of a bipartite entangled state in the presence of ADC in a photonic system. The polarization entangled photons are sent to two displaced Sagnac interferometers where ADC is simulated by rotating HWP $H_1$ and the NOT operation is performed by HWP $H_5$, and then ADC is continued by a set of secondary HWPs $H_2~\&~H_6$. The path lengths of the photons reaching in mode $|b\rangle$ are compensated for coherent recombination of polarization amplitudes at PBS $P_1$. This ensures that when all the ADC HWPs are set to zero, except for NOT operation, and an initial entangled state $|\alpha||HH\rangle+|\beta|\exp(\iota\delta)|VV\rangle$ is incident at the input ports of the interferometers, we reconstruct the state $|\alpha||VV\rangle+|\beta|\exp(\iota\delta)|HH\rangle$ at the output ports. These photons are finally sent for quantum state tomography (QST). The HWP $H_4$ is used to flip the polarization of photons passing through it such that the QST settings remain the same for the photons in all spatial modes.}}
\end{center}
\end{figure}

The initial state of the system (5) in the presence of ADC (due to $H_1$ at $\theta$)  evolves as follows,

\begin{equation}
\rho^{(2)}(p,0)=\sum _{i,j} M_{ij}~\rho(0,0)~M_{ij}^\dag~~;~~i,j=1,2 .
\end{equation} 

Apply the NOT operation on both the qubits at $p=p_n$ as follows,

\begin{equation}
\rho^{(2)}(p_n,0)=(\hat{\sigma_x} \otimes \hat{\sigma_x}) \rho^{(2)}(p,0)~(\hat{\sigma_x} \otimes \hat{\sigma_x})^\dag ,
\end{equation}
where $\hat{\sigma_x}$ is the Pauli matrix. This amounts to switching the elements $\rho_{11}$ and  $\rho_{44}$ and  $\rho_{22}$ and  $\rho_{33}$ and interchanging (complex conjugation) the off-diagonal elements.

Apply the Kraus operators $M'_{ij}$ to complete the evolution of the system in the presence of ADC (due to $H_2$ and $H_6$ at $\theta'$) after the NOT operation as follows,

\begin{equation}
\rho^{(2)}(p_n,p')=\sum _{i,j} M'_{ij}~\rho^{(2)}(p_n,0)~M_{ij}^{\prime\dag}~~;~~i,j=1,2, 
\end{equation}

with entries now in the form of (7) given by,

\begin{equation}
\begin{aligned}
\rho^{(2)}_{11}(p_n,p')&=(1 - p_n)^2 x + 2 p' (1 - p_n) p_n x + p'^2 (u + p_n^2 x),\\
\rho^{(2)}_{22}(p_n,p')&=(1 - p') (1 - p_n) p_n x + (1 - p') p' (u + p_n^2 x),\\
\rho^{(2)}_{33}(p_n,p')&=(1 - p') (1 - p_n) p_n x + (1 - p') p' (u + p_n^2 x),\\
\rho^{(2)}_{44}(p_n,p')&=(1 - p')^2 (u + p_n^2 x),\\
\rho^{(2)}_{14}(p_n,p')&=(1 - p') (1 - p_n) v^*,\\
\rho^{(2)}_{41}(p_n,p')&=(1 - p') (1 - p_n) v.\\    
\end{aligned}
\end{equation}

\section{Effect of the NOT operation applied on only one of the qubits}

The proposed experimental set up for studying the effect of a NOT operation applied on only one of the qubits of a bipartite entangled state in the presence of ADC on the dynamics of entanglement is to retain one half, say the lower, as in figure (1) and have only the upper half as in figure (2), the optical elements $H_5$ and $H_6$ occurring only in the upper arm.

The initial state of the system (5) in the presence of ADC (due to $H_1$ at $\theta$)  evolves as follows,

\begin{equation}
\rho^{(3)}(p,0)=\sum _{i,j} M_{ij}~\rho(0,0)~M_{ij}^\dag~~;~~i,j=1,2.
\end{equation} 

Apply the NOT operation on only one of the qubits by $H_5$ at $45^o$, let us say first qubit, at $p=p_n$ as follows,

\begin{equation}
\rho^{(3)}(p_n,0)=(\hat{\sigma_x} \otimes \hat{\mathbb{I}})~ \rho^{(3)}(p,0)~(\hat{\sigma_x} \otimes \hat{\mathbb{I}})^\dag .
\end{equation}

Apply next the Kraus operators $M'_{ij}$ to complete the evolution of the system in the presence of ADC (due to $H_2~\&~H_6$ at $\theta'$) after the NOT operation to give,

\begin{equation}
\rho^{(3)}(p_n,p')=\sum _{i,j} M'_{ij}~\rho^3(p_n,0)~M_{ij}^{\prime\dag}~~;~~i,j=1,2,
\end{equation}

with entries now in the form of (7) given by,

\begin{equation}
\begin{aligned}
\rho^{(3)}_{11}(p_n,p')&= p' (1 - p_n)^2 x + (1 - p_n) p_n x + p'^2 (1 - p_n) p_n x + p' (u + p_n^2 x),\\
\rho^{(3)}_{22}(p_n,p')&=(1 - p') (1 - p_n)^2 x + (1 - p') p' (1 - p_n) p_n x,\\
\rho^{(3)}_{33}(p_n,p')&=(1 - p') p' (1 - p_n) p_n x + (1 - p') (u + p_n^2 x),\\
\rho^{(3)}_{44}(p_n,p')&=(1 - p')^2 (1 - p_n) p_n x,\\
\rho^{(3)}_{23}(p_n,p')&=(1 - p') (1 - p_n) v^*,\\
\rho^{(3)}_{32}(p_n,p')&=(1 - p') (1 - p_n) v.\\    
\end{aligned}
\end{equation}

\section{Some analytical expressions}
 
Let the two polarization entangled qubits constitute the system, as given by eqn (5), and the action of the rotating HWPs simulate the ADC. This causes a $|V\rangle$ polarized photon to probabilistically ``decay`` to $|H\rangle$ with probability  $p=sin^2(2\theta)$ ($p'=sin^2(2\theta')$), where $\theta$ ($\theta'$) is the angle between the fast axis of the HWP and $|V\rangle$. The ADC probability $p'_0$ at which ESD happens, depends on the initial state parameters of the entangled system and the ADC setting of first HWP $p$. The criterion for ESD as indicated by a switch in sign of the eigenvalues of the partial transpose of Eq(7) is given by $\rho_{22}\rho_{33}=|\rho_{14}|^2$. For the initial state (5), the condition for ESD is obtained by computing the Negativity of the state (7) and equating it to zero. The condition for ESD is given by, 

\begin{equation}
p'_0=\frac{|v|-xp}{ x(1-p)}.
\end{equation}

Let us denote the effective end of entanglement due to combined evolution through two HWPs by $p_{end}$. The $p_{end}$ involves a multiplication of survival probabilities to give,
\begin{equation}
1-p_{end}=(1-p)(1-p'_0)~~ \text{with}~~ p_{end} =|v|/x.
\end{equation}
depending only on the initial state parameters in (5).

For the manipulation of ESD using NOT operation on both the qubits, this operation switches $\rho_{11}$ and  $\rho_{44}$, $\rho_{22}$ and  $\rho_{33}$  , and interchanges the off-diagonal elements $\rho_{14}$ and $\rho_{41}$ in Eq. (7). With subsequent evolution, the criterion for when ESD now happens, can be used to determine the value of $p_A$ that marks the boundary between hastening or not relative to $p_{end}$, and similarly the value $p_B$ that is the boundary between delaying $p_{end}$ past $p_0$ or averting ESD completely. We get,
\begin{equation}
p_A=\frac{1-2 u}{2 (1-u)}~~,~~p_B=\frac{|v|-u}{1+|v|-u}.
\end{equation}

For the manipulation of ESD using NOT operation on only one of the qubits, this operation now switches $\rho_{11}$ and $\rho_{33}$ , $\rho_{22}$  and $\rho_{44}$ , and moves $\rho_{14}$  into the $\rho_{23}$  position. Following subsequent evolution, the ESD criterion through the partial transpose matrix now becomes $\rho_{11} \rho_{44}=|\rho_{23}|^2$. We now get,

\begin{equation}
p_A=\frac{|v|}{u+ 2|v|}~~,~~p_B=\frac{|v|^2}{|v|^2-u+1}.
\end{equation}

These simple expressions defining the time $p_0$ for ESD, and the times for NOT operation that define the delay/hasten and avert/delay boundaries may also be given for a more general mixed state density matrix with also non-zero entries in the two other diagonal position in (5) and are recorded in Appendix [B]. The Appendix [C] records similar expressions for a density matrix with non-zero values in the other off-diagonal position as in [35].

The NOT operation applied on both the qubits at $p=p_n$ of a bipartite entangled state leads to the end of entanglement given by,

\begin{equation}
p_{end}=\frac{p_n^2 (2x+|v|)+p_n (1-2x-2|v|)+|v|}{x \left(p_n^2-1\right)+1}.
\end{equation}

When NOT operation is applied on only one of  the qubits at $p=p_n$, the end of entanglement is given by,

\begin{equation}
p_{end}=\frac{[4 x (p_n-1) p_n + 4 (p_n-1)^2 |v|^2+1]^{1/2}+2 x p_n-1}{2x p_n}.
\end{equation}

\section{Results and Discussion} 

As an example, we choose the initial state $|\psi\rangle=|\alpha||HH\rangle+|\beta|  \exp{(\iota\delta)}|VV\rangle$ with $|\alpha|=1/\sqrt{5},|\beta|=2/\sqrt{5}$ and $\delta=0$ and report the results for ESD, and ESD-manipulation using the NOT operation on one or both the qubits of the bipartite entangled state.

For ESD using two HWPs, the  disentanglement happens for $p=0$ at $p'_0=0.5$  and for any other combination of $p$ and $p'$, $p'_0$ follows the non-linear eqn (17) in $p$, and the effective end due to two HWPs is given by non-linear eqn (18) with $p_{end}=0.5$. The plot of Negativity N vs. probability of decay of qubits ($p, p'$) for the state (7) is shown in figure (3). The plot of Purity (defined as $Tr(\rho^2)$) vs. probability of decay of qubits ($p,p'$) for ESD is shown in figure (4). The two qubit entangled state (5), initially in a pure state, gets mixed at intermediate stages of amplitude damping and finally becomes pure again when both the qubits have decohered  down to the ground state ($|HH\rangle$) at $p=1$ or $p'=1$. However, at ESD for $p_0=0.5$, it ends as a mixed disentangled state.

\begin{figure}[h]
\begin{center}
\includegraphics[scale=0.4]{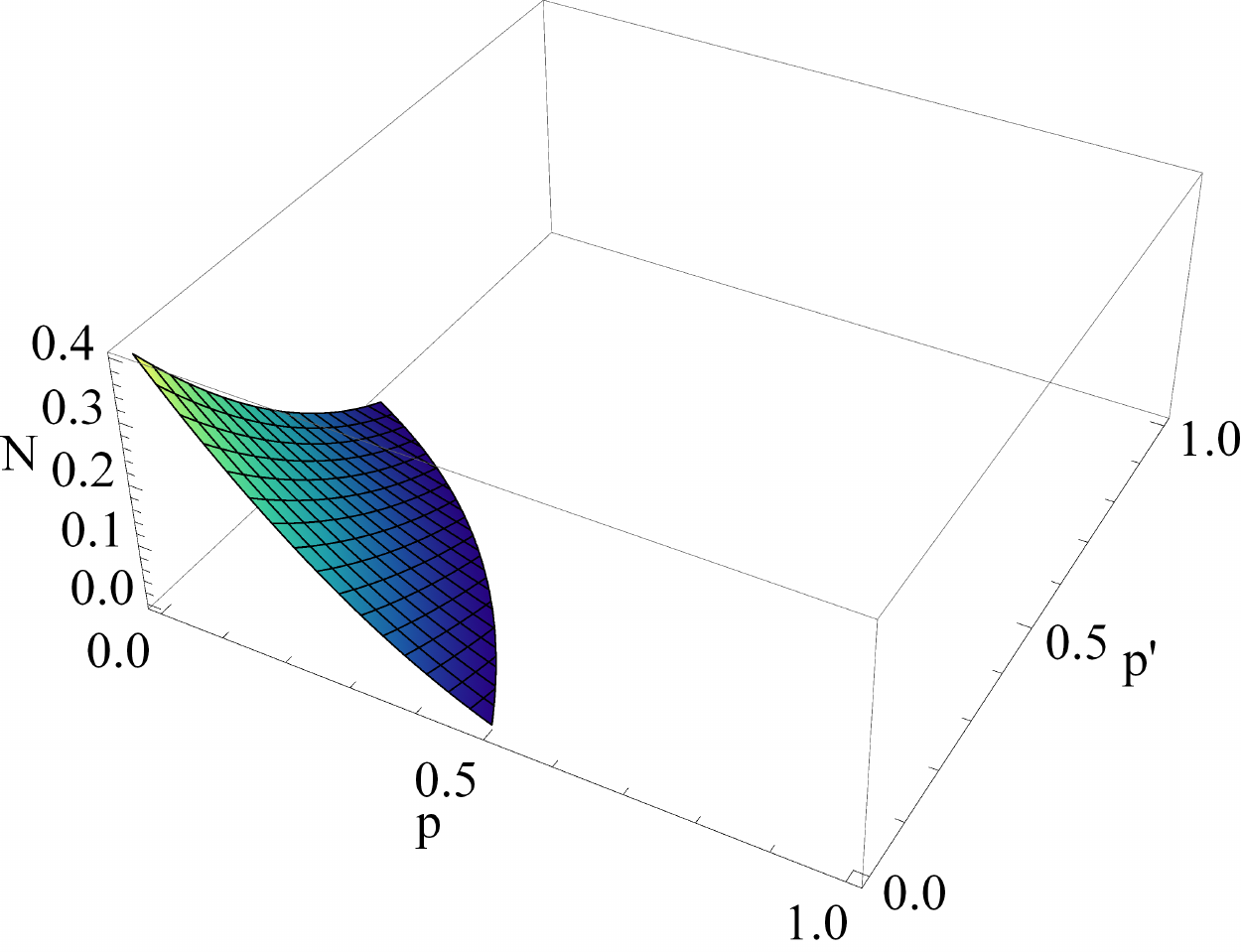}  
\caption{\textit{Plot of Negativity N vs. ADC probability ($p,p'$) for ESD using two HWPs for a bipartite entangled state. The plot implies ESD for $p=0,~p'=0.5$ and $p'=0,~ p=0.5$. For intermediate values of $p~(p')$, the curvature reflects the non-linear relation between them as per (17), the action of two HWPs oriented at $\theta$ and $\theta'$ applied one after another is not equivalent to that of one HWP oriented at $\theta+\theta'$.}}
\end{center}
\end{figure}

 \begin{figure}[h]
\begin{center}
\includegraphics[scale=0.4]{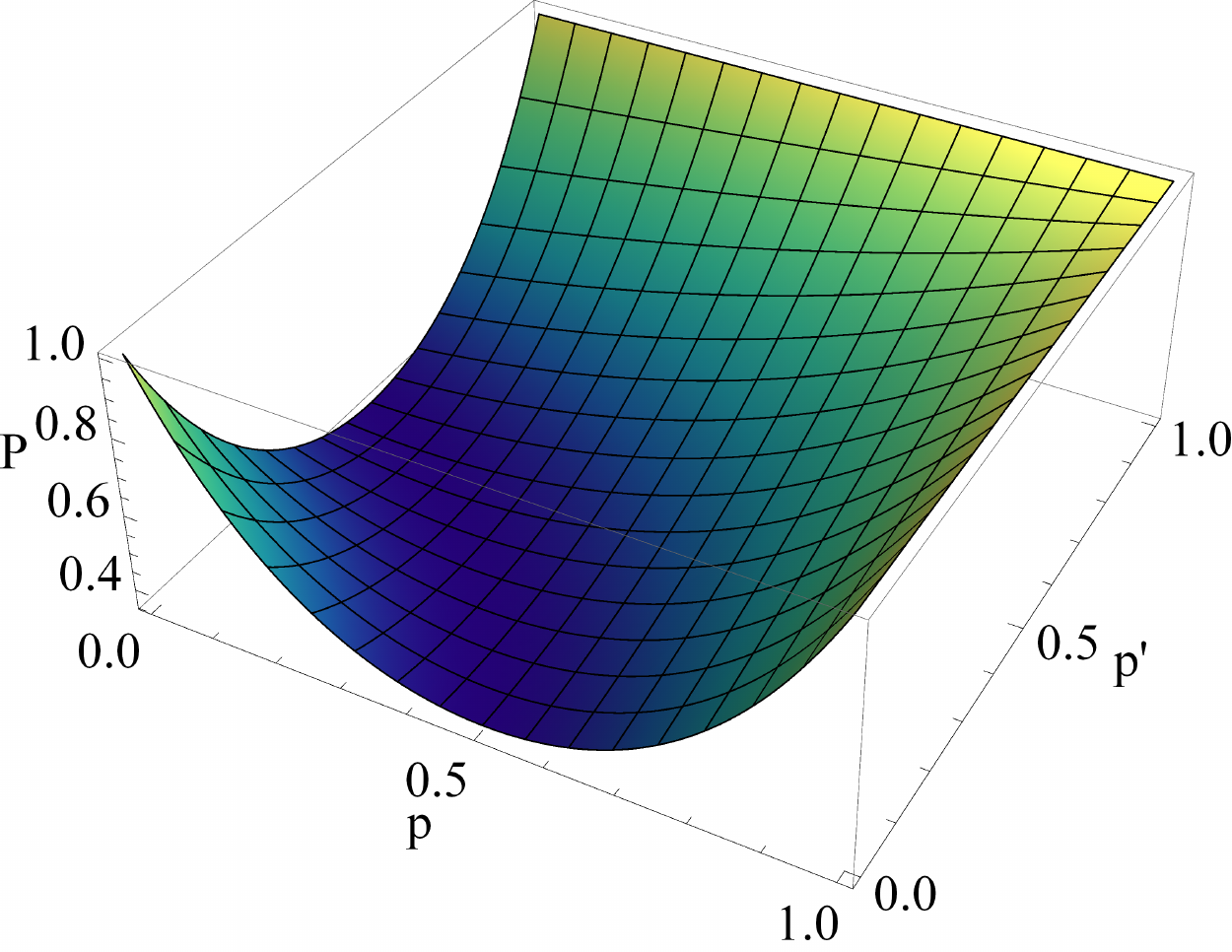} 
\caption{\textit{Plot of Purity P vs. ADC probability ($p,p'$) for ESD using two HWPs for a bipartite entangled state. The state (7) is initially pure, gets mixed at intermediate ADC probabilities and becomes pure again at $p=1$ or $p'=1$.}}
\end{center}
\end{figure}

For the manipulation of ESD using NOT operation on both the qubits: we get $p_A=0.375$, and $p_B=0.1667$. The corresponding plot of Negativity $N$ vs. ADC probability ($p_n , p'$) for the state (11) is  shown in figure (5). For the  manipulation of ESD using NOT operation on only one of the qubits: we get $p_A=0.4,p_B=0.1667$. The corresponding plot of Negativity $N$ vs. ADC probability ($p_n , p'$) for the state (15) is  shown in figure (6).

 \begin{figure}[h]
\begin{center}
\includegraphics[scale=0.4]{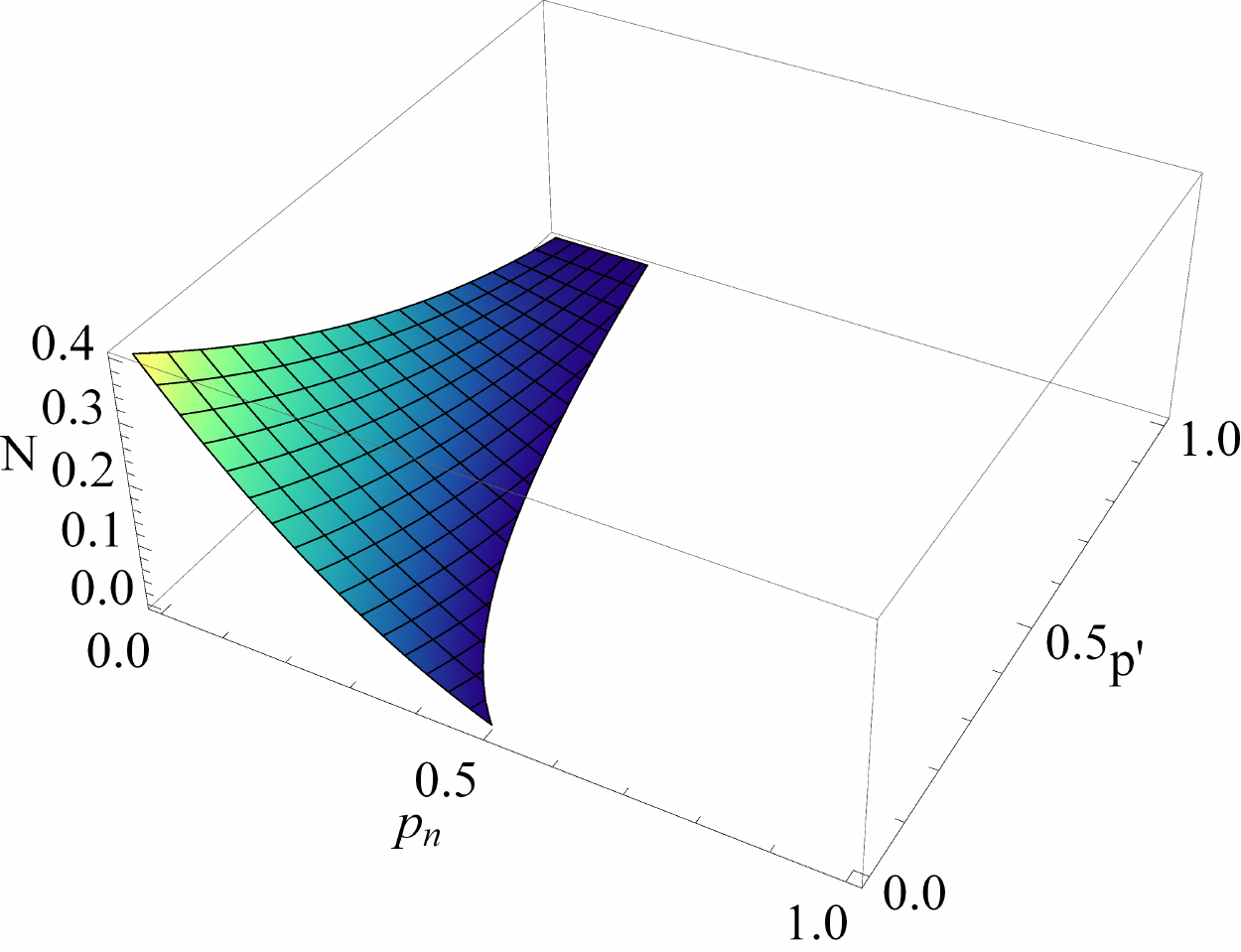}
\caption{\textit{Plot of Negativity N vs. ADC probability ($p_n,p'$) such that NOT operation is applied on both the qubits at $p=p_n$ for manipulation of ESD of a bipartite entangled state. The NOT operation leads to hastening for $0.375 < p_n < 0.5$, delay for $0.1667 < p_n < 0.375$, and  avoidance of ESD for $0\leq p_n \leq 0.1667$.}}
\end{center}
\end{figure}

\begin{figure}[h]
\begin{center}
\includegraphics[scale=0.4]{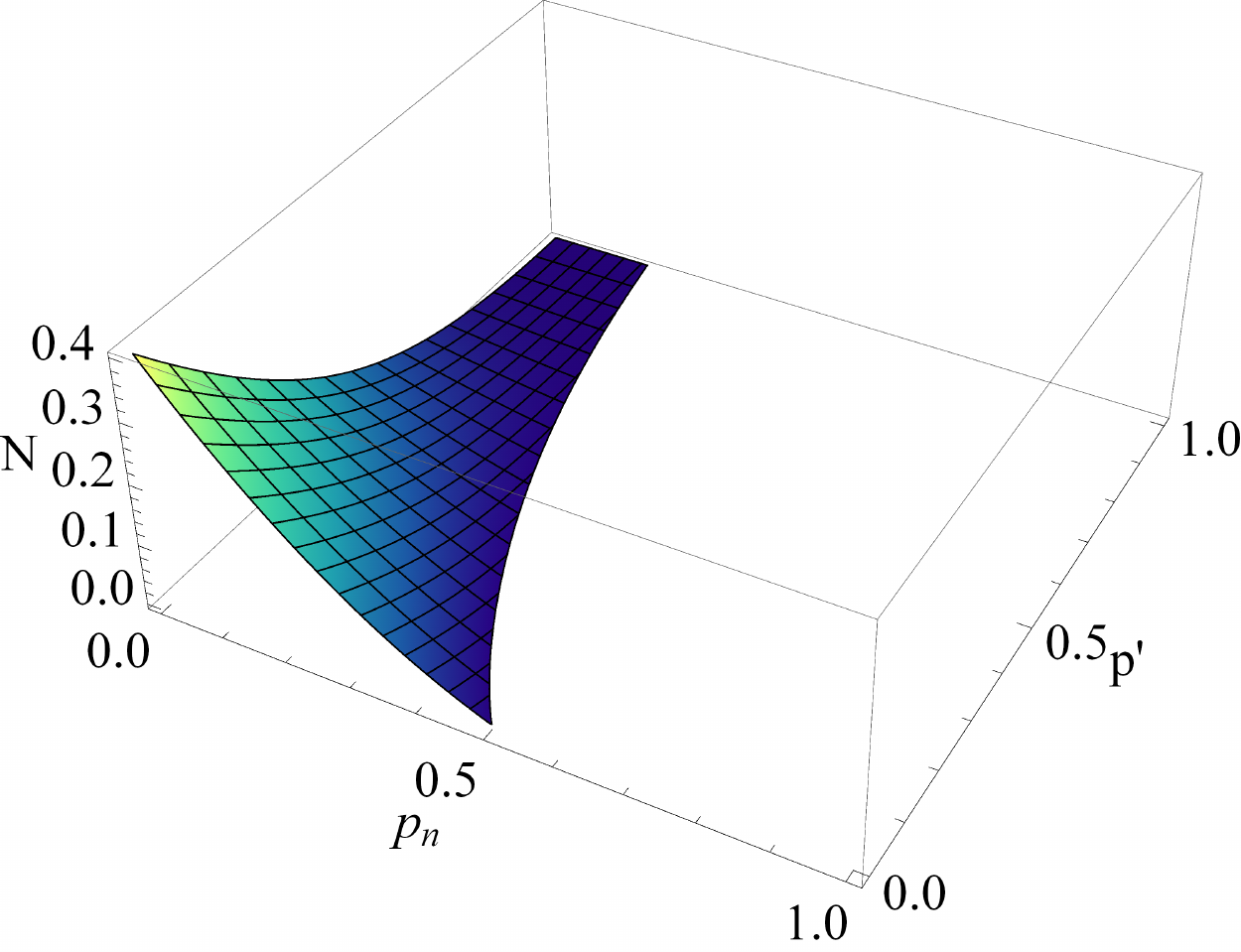} 
\caption{\textit{Plot of Negativity N vs. ADC probability ($p_n,p'$) such that NOT operation is applied on only one of the qubits at $p=p_n$ for manipulation of ESD of a bipartite entangled state. The NOT operation leads to hastening for $0.4 < p_n <0.5$, delay for $0.1667 < p_n < 0.4$, and avoidance of ESD for $0\leq p_n \leq 0.1667$.}}
\end{center}
\end{figure}

 The plot of $p_{end}$ vs. $p_n$ for eqn (21) and (22) such that the NOT operations applied on both (only one of) the qubits at $p_n$ leads to disentanglement at $p_{end}$ is shown by dashed (solid) blue curve in figure (7). In the avoidance range $0 \leq p_n \leq 0.1667$, the $p_{end}$ vs. $p_n$ curves are cut off at $p_{end}=1$ to signify the asymptotic decay with probabilities remaining in the physical domain. For comparison, we have also included the results of ESD; eqn (17) and a rendering of (18), for every $p$, giving the value of $p'_0$, the compounding of them giving the flat line at  $p_{end}=0.5$ as shown by dotted red curve and dot-dashed red line in figure (7). The role of NOT operation on manipulation of ESD is evident as for (i) $0 \leq  p_n \leq 0.1667$, we get avoidance of ESD with $p_{end}=1$ in this range for NOT operation on only one or both the qubits (ii) $0.1667 < p_n < 0.375$ ( $0.1667 < p_n < 0.4$), we get delay of ESD as the dashed (solid) blue curve lies above 0.5 but less than 1, and (iii) $0.375 < p_n < 0.5$ ($0.4 < p_n < 0.5$), we get hastening of ESD as the dashed (solid) blue curve dips below 0.5.

\begin{figure}[h!]
\centering 
\includegraphics[scale=0.75]{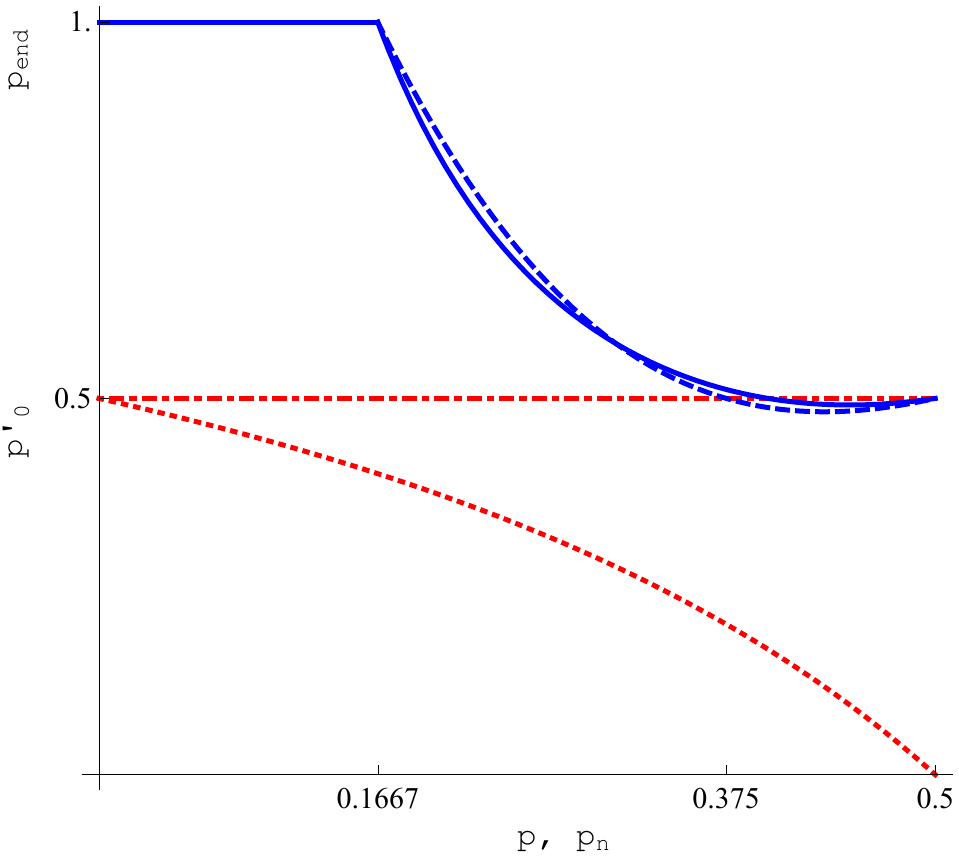} 
\caption {\textit{ Plot shows the effect of NOT operation on manipulation of ESD at various $p_n$ values for density matrix parameters $u=0.2, |v|=0.4$. The plot for ESD, eqn (17) and a rendering of (18), for every $p$, giving the value of $p'_0$, the compounding of them giving the flat line $p_{end}=0.5$  are shown by dotted red curve and dot-dashed red line. The plot for manipulation of ESD using NOT on both or only one qubit are shown by dashed and solid blue curves, respectively. For $0\leq p_n \leq 0.1667$, $p_{end}=1$ means avoidance of ESD, the $p_B$ value for single and double NOT coinciding for these particular parameters but $p_B$ is in general different. For $0.1667 < p_n < 0.4$ ( $0.1667 < p_n < 0.375$) the dashed (solid) blue curve lies above 0.5, implies delays of ESD for double (single) NOT manipulation of ESD. The dashed (solid) blue curve dipping below 0.5 for $0.375 < p_n < 0.5$ ($0.4 < p_n < 0.5$) for manipulation using double (single) NOT operation implies hastening of ESD in this range.}}
\end{figure}

The discussion so far, and figures $3-7$, pertain to the choice $u=0.2, |v|=0.4$, and result in $p_0=0.5, p_B=0.1667, p_A=0.375$ for NOT applied to both whereas $p_A=0.4$ when applied to just one qubit. This is an example when $p_A > p_B$ and both lie in the physically relevant interval ($0, p_0$). All three phenomena, of hastening ($p_A < p_{NOT} < p_0$), delaying ($p_B < p_{NOT} < p_A$), and averting ($0 \leq p_{NOT}\leq p_B$) ESD then occur. The appearance of the various manipulation regimes (hastening, delay, and avoidance of ESD) critically depends on the choice of the parameters of the initial state (density matrix) of the system as expressed in eqn (17-20).

Consider a general initial state (5), with $|v|\leq \sqrt{u(1-u)}$, which captures pure as well as mixed entangled states. The condition for the existence of hastening regime is $u+|v| > 0.5$ for manipulation of ESD using single or double NOT operation. The condition for existence of avoidance regime is $|v|>0$ ($|v|>u$) for manipulation of ESD using single (double) NOT operation. For the pure entangled state (4), the condition for a physically relevant $p_A$ is $u\geq 0.1464$. Thus, pure entangled states (4) with $0.1464\leq u <0.5$ give rise to hastening, delay as well as avoidance of ESD, whereas states with $0<u < 0.1464$ give rise to delay and avoidance of ESD only. For all values of initial parameters, the analytical expressions in (17-20) provide $p_0$ for ESD, and $p_A$ and $p_B$. When these lie within the domain $(0,p_0)$, all three regimes are realized. Otherwise, one may have only two or one of the three regimes of avoidance, delay or hastening of ESD. More general expressions for a wider class of density matrices than (5) are  given in Appendix [B,C]. 

Consider another example of pure entangled state of the form (5) with $u=0.14$ and $|v|=0.347$. For this state, $p_0=0.4035$,  $p_B=0.1228$ ($p_B=0.1715$) and $p_A$ does not exist in the physical domain for single (double) NOT operation. Therefore, we get only delay and avoidance of ESD. The corresponding plot of $p_{end}$ vs. $p_n$ is shown by solid (dashed) blue curve in figure(8). Next, consider an example of mixed entangled state of the form (5) with $u=0.2$ and $|v|=0.15$. For this state, $p_0=0.1875$, $p_B=0.0274$ ($p_B$ does not exist), and  $p_A$ does not exist in the physical domain for single (double) NOT operation. Therefore, NOT operation applied on only one (both) of the qubits delays as well avoids (only delays) the ESD. The corresponding plot of $p_{end}$ vs. $p_n$ is shown by solid (dashed) blue curve in figure (9).

\begin{figure}[h!]
\begin{center}
\includegraphics[scale=0.85]{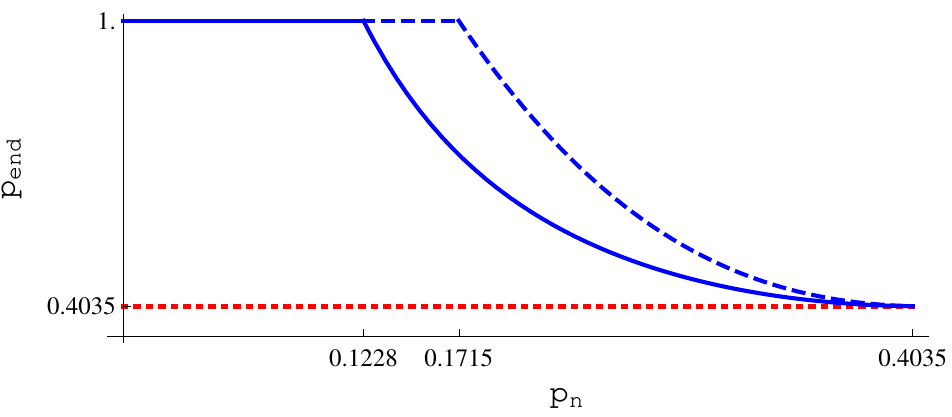} 
\caption {\textit{Plot of $p_{end}$ vs. $p_n$ for manipulation of ESD using NOT operation on one (solid blue curve) or both (dashed blue curve) the qubits of a bipartite entangled state in the presence of ADC for density matrix parameters $u=0.14, |v|=0.347$. The NOT operation leads to avoidance for $0 \leq p_n \leq 0.1228$  ($0 \leq p_n \leq 0.1715$) with disentanglement happening at $p_{end}=1$ in this range, and  delay of ESD for $0.1228 <  p_n < 0.4035$ ($0.1715 < p_n < 0.4035$) as $p_{end}$ curve lies above the $p_{end}=0.4035$ dotted red ESD line for single (double) NOT operation. There is no hastening of ESD for this particular choice of parameters for single or double NOT operation.}}
\end{center}
\end{figure}

\begin{figure}[h!]
\begin{center}
\includegraphics[scale=0.85]{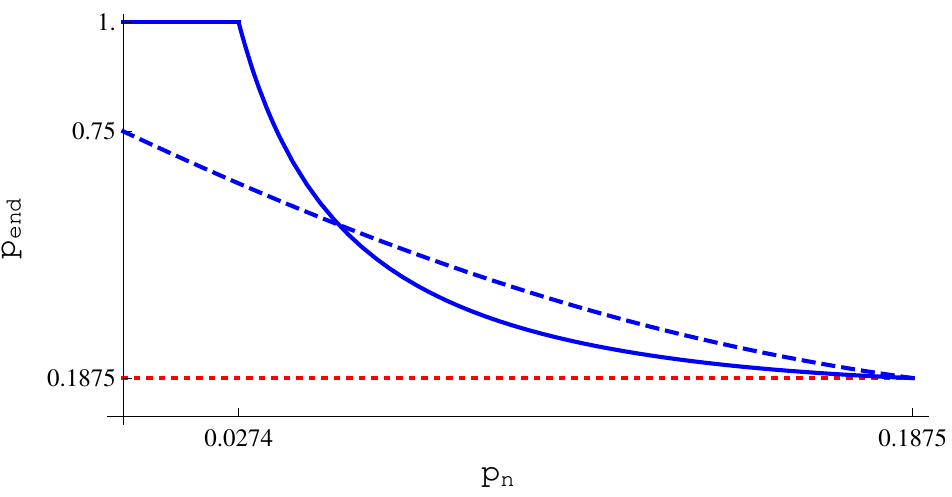} 
\caption {\textit{Plot of $p_{end}$ vs. $p_n$ for manipulation of ESD using NOT operation on one (solid blue curve) or both (dashed blue curve) the qubits of a bipartite entangled state in the presence of ADC for density matrix parameters $u=0.2, |v|=0.15$. The NOT operation when applied on only one of the qubits leads to avoidance of ESD for $ 0 < p_n < 0.0274$  with disentanglement happening at $p_{end}=1$ in this range, and delay for $0.0274 < p_n < 0.1875$ as the  solid blue curve lies above the $p_{end}=0.1875$ dotted red ESD line. The NOT operation when applied on both the qubits leads to delay for $0 < p_n < 0.1875$ as the dashed blue curve lies above the $p_{end}=0.1875$ dotted red ESD line. There is no hastening of ESD for single or double NOT operation and no avoidance of ESD for double NOT operation for this particular choice of parameters.}}
\end{center}
\end{figure}

\section{Conclusions and future work}

We have proposed an all-optical experimental setup for the demonstration of hastening, delay, and avoidance of ESD in the presence of ADC in a photonic system. The simulation results of the manipulation of ESD considering a photonic system, when NOT operations are applied on one or both the qubits, are completely consistent with the theoretical predictions of reference [35] for the two-level atomic system where spontaneous emission is the ADC. We give analytical expressions for $p_0~,p_A~\&~p_B$ which depend on the parameters of the density matrix of the system for both the forms considered here in (5) and that in [35].

Our proposal also has an advantage over decoherence suppression using weak measurement and quantum measurement reversal, and delayed choice decoherence suppression. There, as the strength of weak interaction increases, the success probability of decoherence suppression decreases. In our scheme, however, we can manipulate the ESD, in principle, with unit success probability as long as we perform the NOT operation at the appropriate wave plate angle which is analogous to time in the atomic system. Delay and avoidance of ESD, in particular, will find application in the practical realization of quantum information and computation protocols which might otherwise suffer a short lifetime of entanglement. Also, it will have implications towards such control over other physical systems. The advantage of the manipulation of ESD in a photonic system is that one has complete control over the damping parameters, unlike in most atomic systems. An experimental realization of our proposal will be important for practical noise engineering in quantum information processing, and is under way. Further work in the future could study the dynamics of entanglement in the presence of the generalized ADC [40-43] and the squeezed generalized ADC [44] and the possible schemes for manipulation of entanglement sudden death in the presence of such damping channels.

\section{Acknowledgments}
US and AS acknowledge Prof. Ujjwal Sen for his comments on the manuscript. AS acknowledges Subhajit Bhar for his assistance in literature review as well as verification of some calculation steps. ARPR thanks the Raman Research Institute for its hospitality during visits.

\section{References}

\begin{enumerate}
\item \textit{E. Schr$\ddot{o}$dinger, "Naturwissenschaften" \textbf{23}, 807 (1935)}. 
\item \textit{A. Einstein, B. Podolsky, N. Rosen, "Can the Quantum-mechanical description of a physical reality be considered complete?" Phys. Rev. \textbf{47}, 777 (1935).} 
\item \textit{ J. S. Bell, "On the Einstein Podolsky Rosen paradox", Physics (Long Island City, N.Y.) \textbf{1}, 195-200 (1964).}
\item \textit{S. J. Freedman and J. F. Clauser, "Experimental Test of Local Hidden-Variable Theories", Phys. Rev. Lett. \textbf{28}, 938 (1972).}
\item \textit{T. Yu  and J. H. Eberly , "Finite-Time Disentanglement Via Spontaneous Emission", Phys. Rev. Lett. \textbf{93}, 140404 (2004).}
\item \textit{T. Yu  and J. H. Eberly , "Quantum Open System Theory: Bipartite Aspects", Phys.Rev. Lett. \textbf{97}, 140403 (2006).}
\item \textit{T. Yu  and J. H. Eberly , "Sudden Death of Entanglement", Science \textbf{ 323} , 598 (2009).}
\item \textit{J. Laurat, K. S. Choi, H. Deng, C. W. Chou, H. J. Kimble, "Heralded Entanglement between Atomic Ensembles: Preparation", Decoherence, and Scaling,  Phys. Rev. Lett. \textbf{99}, 180504 (2007).}
\item \textit{M. P. Almeida, F. de Melo, M. Hor-Meyll, A. Salles, S. P. Walborn,
P. H. Souto Ribeiro, L. Davidovich, "Environment-Induced Sudden Death of Entanglement", Science \textbf{316}, 555 (2007).}
\item \textit{Jin-Shi Xu, Chuan-Feng Li, Ming Gong, Xu-Bo Zou, Cheng-Hao Shi, Geng Chen, and Guang-Can Guo, "Experimental demonstration of photonic entanglement collapse and revival", Phys. Rev. Lett.\textbf{ 104}, 100502 (2010).}
\item \textit{R. Horodecki, P. Horodecki, M. Horodecki, K. Horodecki, "Quantum entanglement",  Rev. Mod. Phys.\textbf{ 81}, 865 (2009).}
\item \textit{C. H. Bennett and D. P. DiVincenzo, “Quantum information and computation,” Nature \textbf{404}, 247 (2000).}
\item \textit{M. A. Nielsen  and I. L. Chuang, "Quantum Computation and Quantum Information" (Cambridge University Press, Cambridge, 2000).}
\item \textit{J.W. Pan, S. Gasparoni, R. Ursin, G. Weihs, \& A. Zeilinger,  "Experimental entanglement purification of arbitrary unknown states", Nature \textbf{423},
417 (2003).}
\item\textit{P. G. Kwiat, S. Barraza-Lopez, A. Stefanov, and N. Gisin, “Experimental entanglement distillation and ‘hidden’ non-locality,” Nature \textbf{ 409}, 1014 (2001).}
\item \textit{D. A. Lidar, I. L. Chuang, and K. B. Whaley, “Decoherence-free subspaces for quantum computation,” Phys. Rev. Lett.\textbf{ 81}, 2594 (1998).}
\item \textit{P. G. Kwiat, A. J. Berglund, J. B. Altepeter, and A. G. White, “Experimental verification of decoherence-free subspaces,” Science\textbf{ 290}, 498 (2000).}
\item \textit{D. Kielpinski, V. Meyer, M. A. Rowe, C. A. Sackett, W. M. Itano, C. Monroe, and D. J. Wineland, "A decoherence-free quantum memory using trapped ions", Science\textbf{ 291}, 1013 (2001).}
\item \textit{L. Viola, E. M. Fortunato, M. A. Pravia, E. Knill, R. Laflamme, and D. G. Cory, "Experimental realization of noiseless subsystems for quantum information processing", Science \textbf{293}, 2059 (2001).}
\item \textit{P.W. Shor, "Scheme for reducing decoherence in quantum computer memory", Phys. Rev. A \textbf{52}, R2493  (1995) .}
\item \textit{A.M. Steane," Error Correcting Codes in Quantum Theory", Phys. Rev. Lett. \textbf{77} , 793 (1996) .} 
\item \textit{Lorenza Viola, Emanuel Knill and Seth Lloyd, "Dynamical Decoupling of Open Quantum Systems", Phys. Rev. Lett. \textbf{82}, 2417 (1999).}
\item \textit{Michael J. Biercuk, Hermann Uys, Aaron P. VanDevender, Nobuyasu Shiga, Wayne M. Itano \& John J. Bollinger, "Optimized dynamical decoupling in a model quantum memory", Nature \textbf{458}, 996 (2009).}
\item\textit{Jiangfeng Du, Xing Rong, Nan Zhao, Ya Wang, Jiahui Yang \& R. B. Liu, "Preserving electron spin coherence in solids by optimal dynamical decoupling" , Nature \textbf{461}, 1265 (2009).}
\item \textit{P. Facchi, D. A. Lidar, and S. Pascazio, "Unification of dynamical decoupling and the quantum Zeno effect", Phys. Rev. A \textbf{69}, 032314 (2004).}
\item \textit{ S. Maniscalco, F. Francica, R. L. Zaffino, N. L. Gullo, and F. Plastina, "Protecting entanglement via the quantum Zeno effect", Phys. Rev. Lett. \textbf{100}, 090503 (2008).}
\item \textit{J. G. Oliveira, Jr., R. Rossi, Jr., and M. C. Nemes, “Protecting, enhancing, and reviving entanglement,” Phys. Rev. A \textbf{78}, 044301 (2008).}
\item \textit{Y.S. Kim, Y.W. Cho, Y.-S. Ra, and Y.-H. Kim, "Reversing the weak quantum measurement for a photonic qubit", Opt. Express\textbf{ 17}, 11978 (2009).}
\item \textit{A. N. Korotkov and K. Keane, “Decoherence suppression by quantum measurement reversal,” Phys. Rev. A \textbf{81},040103(R) (2010).}
\item \textit{ J.C. Lee, Y.C. Jeong, Y.S. Kim, and Y.H. Kim, “ Experimental demonstration of decoherence suppression via quantum measurement reversal,” Opt. Express \textbf{19}, 16309 (2011).}
\item \textit{Q. Sun, M. Al-Amri, L. Davidovich, and M. S. Zubairy, “Reversing entanglement change by a weak measurement,” Phys. Rev. A \textbf{ 82}, 052323 (2010).}
\item \textit{Y.S. Kim, J.C. Lee, O. Kwon, and Y.-H. Kim, “ Protecting entanglement from decoherence using weak measurement and quantum measurement reversal,” Nature Phys. \textbf{8}, 117 (2012).}
\item \textit{H.T. Lim, J.C. Lee, K.H. Hong, and Y.H. Kim, “ Avoiding entanglement sudden death using single-qubit quantum measurement reversal,” Opt. Express \textbf{22}, 19055 (2014).}
\item \textit{Jong-Chan Lee, Hyang-Tag Lim, Kang-Hee Hong, Youn-Chang Jeong, M.S. Kim \& Yoon-Ho Kim, "Experimental demonstration of delayed-choice decoherence suppression", Nature Communications \textbf{5}, 4522 (2014)}.
\item \textit{A.R.P. Rau, M. Ali, and G. Alber, "Hastening, delaying or averting sudden death of quantum entanglement",  EPL \textbf{82}, 40002 (2008) .}
\item \textit{A. Peres , "Separability Criterion for Density Matrices",  Phys. Rev. Lett. \textbf{77}, 1413 (1996).}
\item \textit{M.Horodecki, P. Horodecki  and R. Horodecki , "Separability of mixed states: necessary and sufficient conditions", Phys. Lett. A \textbf{223}, 1 (1996).}
\item \textit{P.G. Kwiat, Edo Waks, Andrew G. White, Ian Appelbaum, and Philippe H. Eberhard, "Ultra bright source of polarization entangled photons", Phys. Rev. A \textbf{60}, 773-776(R) (1999).}
\item \textit{D. F. V. James, P. G. Kwiat, W. J. Munro, A. G. White, "Measurement of qubits",  Phys. Rev. A \textbf{64}, 052312 (2001).}
\item \textit{Akio Fujiwara, "Estimation of a generalized amplitude-damping channel", Phys. Rev. A \textbf{70}, 012317 (2004)}.
\item \textit{Asma Al-Qasimi and Daniel F. V. James, "Sudden death of entanglement at finite temperature",  Phys. Rev. A \textbf{77},012117 (2008)}. 
\item \textit{M. Ali, A. R. P. Rau, and G. Alber, “Manipulating entanglement sudden death of two-qubit X-states in zero- and finite-temperature reservoirs,”  J. Phys. B: At. Mol. Opt. Phys.\textbf{ 42}, 025501(8)(2009)}.
\item\textit{Mahmood Irtiza Hussain, Rabia Tahira and Manzoor Ikram, "Manipulating the Sudden Death of Entanglement in Two-qubit Atomic Systems", Journal of the Korean Physical Society, \textbf{59}, 2901-2904(2011).}
\item \textit{R. Srikanth and Subhashish Banerjee, "Squeezed generalized amplitude damping channel", Phys. Rev. A \textbf{77}, 012318 (2008)}. 
\end{enumerate}

\section{Appendices}
\begin{appendix}

\section{An alternative approach to analyze the ESD and manipulation experiments}
We provide here an alternative and intuitive approach to analyze the ESD and its manipulation set up by tagging the photon polarization states with the spatial modes of the interferometer upon the action of each of optical component encountered in the photon's path. The evolution of system plus reservoir is represented  by a unitary operator $U_{SR}$. The degrees of freedom of the reservoir can be traced out from $U_{SR}$ to get the Kraus operators which govern the evolution of the system by itself.\\

Consider the experimental set up for ESD as shown in fig(1). An incident $|H\rangle$ polarized photon is transmitted though the PBS $P_1$ and traverses the interferometer in a counter-clockwise direction, returns to $P_1$ and is transmitted into spatial mode $|a\rangle$ of the reservoir. The corresponding quantum map is given by, 

\begin{equation}
\begin{aligned}
U_{SR}|H\rangle_S|a\rangle_R \rightarrow |H\rangle_S|a\rangle_R.\\
\end{aligned}
\end{equation}

An incident  $|V\rangle$ polarized  photon is reflected by PBS $P_1$ and traverses the interferometer in a clockwise direction. The action of HWPs $H_1$ and $H_2$ and  PBS $P_1$ and $P_2$ are represented by the quantum map, 

\begin{equation}
\begin{aligned}
U_{SR}|V\rangle_S|a\rangle_R &\xrightarrow[P_2]{H_1@\theta}\sqrt{1-p} |V\rangle_S|a'\rangle_R+\sqrt{p} |H\rangle_S|b\rangle_R,\\
&\xrightarrow[in~|V\rangle~arm]{H_2@\theta'} \sqrt{1-p}[\sqrt{1-p'}~|V\rangle_S|a'\rangle_R\\& + \sqrt{p'}~ |H\rangle_S|b'\rangle_R] +\sqrt{p}|H\rangle_S|b\rangle_R],\\
\end{aligned}
\end{equation}
where $p=\sin^2(2\theta)~,~p'=\sin^2(2\theta').$

Consider the experimental setup for ESD manipulation using double NOT operation   as shown in figure (2). An incident $|H\rangle$ polarized photon is transmitted through PBS $P_1$ and traverses the interferometer in a counter-clockwise direction where the NOT operation is applied by HWP $H_5$ and ADC afterwards is simulated by $H_6$. The corresponding quantum map is given by,

\begin{equation}
\begin{aligned}
U_{SR}|H\rangle_S|a\rangle_R&\xrightarrow{H_5@45}|V\rangle_S|b\rangle_R~,\\
&\xrightarrow{H_6@\theta'}\sqrt{1-p'}|V\rangle_S|b\rangle_R+\sqrt{p'}|H\rangle_S|a\rangle_R~.
\end{aligned}
\end{equation}

An incident $|V\rangle$ polarized photon is reflected by PBS $P_1$ and traverses the interferometer in a clockwise direction where ADC is introduced by HWP $H_1$ followed by NOT operation by HWP $H_5$ and then ADC is continued by HWP $H_2$. The corresponding quantum map is given by,

\begin{equation}
\begin{aligned}
U_{SR}|V\rangle_S|a\rangle_R &\xrightarrow{H_1@\theta}\sqrt{1-p}|V\rangle_S|a\rangle_R+\sqrt{p}|H\rangle_S|b\rangle_R~,\\
&\xrightarrow[H_5@45]{P_2}\sqrt{1-p}|H\rangle_S|b\rangle_R+\sqrt{p}|V\rangle_S|a'\rangle_R~,\\
&\xrightarrow[in~|V\rangle~arm]{H_2@\theta'}\sqrt{1-p}|H\rangle_S|b\rangle_R\\
&+\sqrt{p}[\sqrt{1-p'}|V\rangle_S|a'\rangle_R
+\sqrt{p'}|H\rangle_S|b'\rangle_R]~.\\
\end{aligned}
\end{equation}

\section{Analytical expressions for X-state with non-zero entries in the other diagonal terms} 

As discussed in Section 5, we again use the negativity criterion for the partial transposed density matrix to determine the occurrence of ESD. By following, as in [35], the evolution of the parameters in Eq.(2), and double NOT switching $a$ and $d$, $b$ and $c$, and swapping the off-diagonal terms $z$ and $z^*$, whereas a single NOT at one end switches $a$ and $c$, $b$ and $d$, and moves $z$ and $z^*$ inward along the anti-diagonal, we again obtain analytical expressions for the various $p$ (or equivalently, $\gamma$ and $t$) of interest. Consider the initial mixed entangled state with the density matrix given in a form more general than (5):
\begin{equation}
\rho_2= \left( \begin{array}{cccc} 
a & 0 & 0 & z\\
0 & b & 0 & 0 \\
0 & 0 & c & 0 \\
z^* & 0 & 0  & d \end{array} \right),
\end{equation}

where $a+b+c+d=1$. As per the convention for ground and excited states in the reference [35], we choose $'a'$ as the population of both qubits being in excited states and $'d'$ being the population of both qubits in the ground states unlike the (reversed) convention in the main body of this paper.

In the presence of ADC, the condition for ESD is given by,

\begin{equation}
p_0=\frac{-b-c+[(b-c)^2+4|z|^2]^{1/2}}{2a}.
\end{equation}

For manipulation of ESD using NOT operation on both the qubits, the condition for hastening ESD is given by,

\begin{equation}
p_A=\frac{a-d}{1+a-d}.
\end{equation}

The condition for avoidance of ESD is given by,

\begin{equation}
p_B =1-\frac{2a+b+c-[(b-c)^2+4|z|^2)]^{1/2}}{2[(a+b)(a+c)-|z|^2]}.\\
\end{equation}

For manipulation of ESD using NOT operation on only one of the qubits, the condition for hastening ESD is given by, 

\begin{equation}
p_A=1-\frac{(c+a)[(c+a)(1-p_0)-1]}{(a+b)\{(a+b)-[(b-c)^2+4|z|^2]^{1/2} \}-a}.
\end{equation}
 
 and the condition for avoidance of ESD is given by,
 
\begin{equation}
p_B=\frac{|z|^2-c}{|z|^2+a}.
\end{equation}

\section{Analytical expressions for the X-state in Reference [35]}

Consider the initial mixed entangled state with the form of density matrix as in reference [35],
\begin{equation}
\rho_1= \left( \begin{array}{cccc} 
a & 0 & 0 & 0 \\
0 & b & z & 0 \\
0 & z^* & c & 0 \\
0 & 0 & 0  & d \end{array} \right).
\end{equation}

For simplicity, the 1/3 factor in [35] has been absorbed into the density matrix elements.\\

In the presence of ADC, the condition for ESD is given by,

\begin{equation}
p_0=\frac{-b-c+[(b+c+2a)^2-4(a-|z|^2)]^{1/2}}{2a},
\end{equation}
where $p=1-\gamma^2=1-\exp{(-\Gamma t)}$, $\Gamma$ is the spontaneous decay rate of the two level atomic qubit as introduced in [35].

For manipulation of ESD using NOT operation on both the qubits, the condition for hastening ESD is given by,

\begin{equation}
p_A=\frac{a-d}{1+a-d}.
\end{equation}

The condition for avoidance of ESD is given by,

\begin{equation}
p_B =\frac{2(a-|z|^2)-(2a+b+c)+[(b+c+2a)^2-4(a-|z|^2)]^{1/2}}{2(a-|z|^2)}.\\
\end{equation}

For manipulation of ESD using NOT operation on only one of the qubits, the condition for hastening ESD is given by,

\begin{equation}
p_A=\frac{(c+a)[2a(1-p_0)-(c+a)(1-p_0)+c+d]-a}{(c+a)[2a(1-p_0)-(b+a)]-a}.
\end{equation}
 
 The condition for avoidance of ESD is given by,
 
\begin{equation}
p_B =1-\frac{a+c}{(a+b)(a+c)+|z|^2}.\\
\end{equation}
 
For hastening, delay and avoidance to exist in a physical region, the corresponding parameters must satisfy the condition $0<p_B~,p_A<p_0$. As an example, the choice $a=0.4, b=c=0.2, z=0.25$ gives $ p_0=0.125$, $p_A=0.1667$, and an unphysical negative value of $p_B$. This means that neither hastening nor averting ESD is possible, only delaying it by applying NOT between $0$ and $p_0$.

\end{appendix}

\end{document}